# The Mpemba Effect in Pure Water Has a Stochastic Origin. Experimental and Theoretical Resolution of the Paradox

Andrei A. Klimov and Alexei V. Finkelstein

Institute of Protein Research, Russian Academy of Sciences, 142290 Pushchino, Moscow, Russia

**The "Mpemba effect" is the name given to the assertion that hot water freezes quicker than cold water[1] or, in a modern and more general form, that the system that is initially more distant from its equilibrium state comes to this state earlier[2]. This counterintuitive statement seems to breach fundamental thermodynamic and kinetic laws; however, numerous experiments[3-10] with classical and quantum systems demonstrate this paradoxical Mpemba effect, leading to extensive discssions in prominent scientific jornals[2,5,9,12-14]. However, the fundamental physical mechanisms behind this effect have remained elusive[14]. Here we performed the water freezing experiments under carefully controlled conditions, and found that the Mpemba effect only occurred when the freezer temperature was very close to the temperature of ice nucleation. In this case, the range of freezing times for both hot and cold water was so great that it exceeded the delayed cooling of the initially hotter liquid, and therefore sometimes the hot water froze before the cold water. Our theoretical analysis of this fact shows that the Mpemba paradox associated with water freezing is rooted in the stochastic nature of ice nucleation, typical of first-order phase transitions. We anticipate our assay to be a starting point for reconsidering the famous Mpemba paradox in water and other systems undergoing similar phase transitions.**

Nearly 60 years ago, the surprising claim that *"hot water freezes faster than cold water"* was dubbed the Mpemba effect[1]. Similar counterintuitive phenomena, seemingly defying basic physics, has since been observed in liquids, polymers, clathrate hydrates, colloids, and even quantum systems This has spurred numerous experimental[3-11,15] and even more theoretical studies[2,9,10,14-30], published, among others, in *Nature* and *Nature Portfolio* journals, highlighting the fact that a seemingly simple process like water freezing was not fully understood...

Nowadays, the initial Mpemba's statement *"hot water freezes* (more precisely: starts to freeze[1]) *faster than cold water"* obtained a more general form "*the state that is initially farthest from its equilibrium state attains the latter at the earliest time*"[2], which increased interest to such phenomena.

While historically linked with sweet milk and water, the Mpemba effect has manifested itself in various systems, sparking increased interest in Mpemba effect-like phenomena; the studies have been extended even to quantum mechanics of many-body systems[2,11,29-30]. However, despite extensive research, the underlying physical mechanisms responsible for this effect remained elusive[14].

To clarify the basis of the in-water Mpemba paradox, we conducted experiments, replicating previously reported water freezing studies[1,3–5,8] under carefully controlled conditions (see Methods). We *sometimes* observed the Mpemba effect – where initially hot water froze faster than initially cold water – but only when the freezer temperature was very close to the ice nucleation temperature.

During our experiments, we noted significant variability in the freezing onset time for both initially hot and cold water (Fig. 1, 2). This variability was most pronounced when the freezer temperature was above -7 °C, and this large spread in freezing onset times *often* (but not always) exceeded the delay in cooling of the initially hotter liquid. However, this "compensation of the delay by variability" only occurred when the freezer temperature was just slightly below the ice nucleation point (Fig. 1). Then, the "stochastic Mpemba effect" was often (in about one third of cases) observed. But when the freezer temperature was well below the ice nucleation point, the range of freezing times was small (Fig. 2), and the Mpemba effect was never observed.

The stochastic nature of the Mpemba effect has been mentioned in a number of studies by cautious phrasing like "the water that starts hotter *may* freeze first"[1] or "hot water *may well begin to freeze* quicker than cold"[3]. However, a thorough study of this critical aspect of the Mpemba effect related to ice nucleation has not yet been conducted. We will now outline and explain the conditions under which the stochasticity in the ice nucleation time can lead to observation of the Mpemba effect.

Consistent with seminal studies[1,3], we focus on the *onset* of freezing, rather than the moment of complete freezing of ice, or the time it takes for water to reach 0 °C. Careful experiments on water cooled to 0°C without freezing[5], demonstrated that hot water *does not* cool faster than cold water. This allows us to exclude factors like water evaporation[31,32] and convection[33,34], and concentrate on the phase transitions, which appear to drive the Mpemba effect in various systems, including milk[1], water[1,3], clathrate hydrates[6], and polymers[7]. It should be noted, though, that Mpemba-like effects can also occur in systems without phase transitions, such as colloids[9,10] and some quantum[2,11,29,30] and mechanical[35] systems.

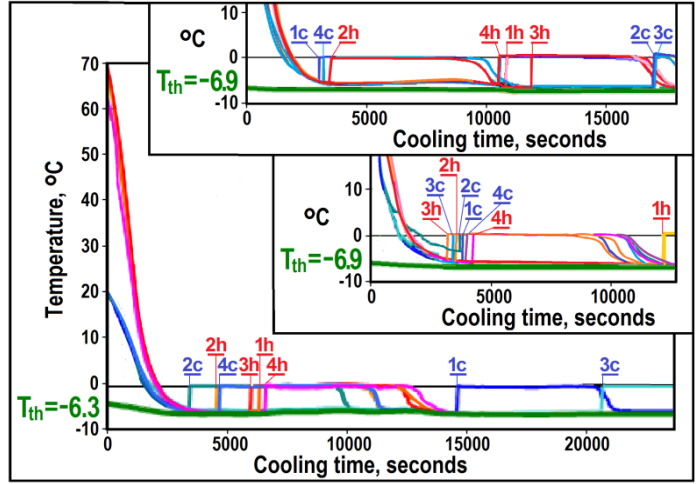

**Fig. 1. Dynamics of water freezing at thermostat temperature $T_{th}$ exceeding -7 °C.** Four water samples (1c – 4c) were initially at +20 °C (blue lines and symbols), and four water samples (1h – 4h) at +70 °C (red lines and symbols). Each line tracks the temperature evolution of one individual sample. The main panel illustrates the full temperature evolution in one experiment. The insets show the low-temperature dynamics in two other experiments. Sharp increases in temperature (marked as 1c – 4c, 1h – 4h) indicate the onset of freezing of the corresponding samples, causing a sharp release of latent heat of freezing[36]. Bold green line represents the measured thermostat temperature $T_{th}$. One can see a large variation in freezing onset times. The ranges of these times *do* overlap for experiments starting at +20 and +70 °C, so that at relatively high (above -7 °C) thermostat temperatures the Mpemba effect is observed for some pairs of samples and not for others (see alternation of blue and red symbols above the plots).

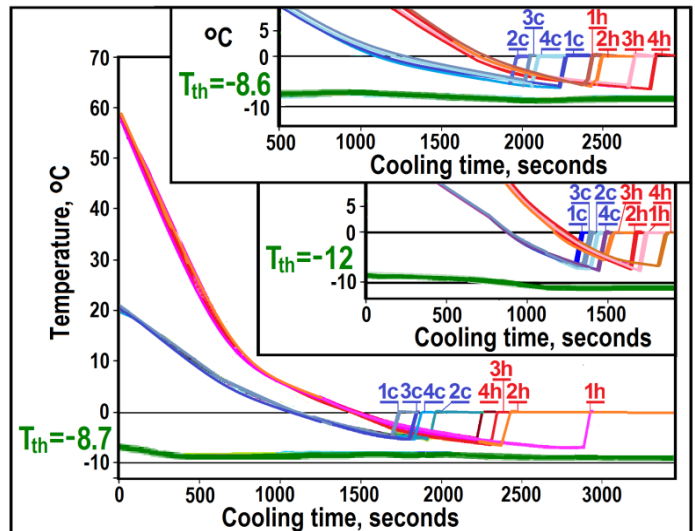

**Fig. 2. Dynamics of water freezing at thermostat temperature $T_{th}$ below -8 °C.** Designations are the same as in Fig. 1. One can see a small variation in freezing onset times. These times *do not* overlap for experiments starting at +20 and +70 °C, so that at relatively low (below -8 °C) thermostat temperatures the Mpemba effect is *not* observed (see separation of blue and red symbols above the plots).

Our water freezing experiments highlight the significant role of freezer temperature in inducing the Mpemba effect. The freezer temperature $T_{\text{th}}$ directly impacts the cooling rate, which, according to Newton's law of thermal conductivity, can be described as

$$dT_t/dt = -\kappa \cdot (T_t - T_{\text{th}}) ; \qquad (1)$$

here $\kappa$ is the thermal conductivity coefficient attributed to the surrounding of the sample (having temperature $T_t$), and $t$ is time. This equation implies that $d[\ln(T_t - T_{\text{th}})] = -\kappa \cdot dt$; so,

$$T_t - T_{\text{th}} = (T_1 - T_{\text{th}}) \cdot \exp(-\kappa \cdot t), \qquad (2)$$

where $T_1 > T_{\text{th}}$ represents the temperature of the sample at the initial time moment ($t_1$=0) of the cooling process. Thus, the time required for the sample to cool from $T_1$ to $T_t$ is

$$t_{\text{cooling}}(T_1, T_t) = \frac{1}{\kappa} \cdot \ln\frac{T_1 - T_{\text{th}}}{T_t - T_{\text{th}}} ; \qquad (3)$$

here, $\kappa \sim 10^{-3}$ s$^{-1}$ corresponds to cooling by a freezer set to $T_{\text{th}} \approx -10$ °C (as in both our experiments and previous works[1,3], where cooling from $T_1 \approx +80$ °C to $T_t \approx +20$ °C took about 20 minutes). Note that in these experiments the difference $T_t - T_{\text{th}}$ decreased 3-fold (from [+80-(-10)]=90 °C to [+20-(-10)]=30 °C) during ~1000 sec, but this "3-fold decrease" can be a fraction of a degree when $T_t$ approaches $T_{\text{th}}$.

The closeness of the thermostat temperature $T_{\text{th}}$ to the ice-nucleation temperature $T_t$ is crucial for the Mpemba effect. This proximity not only enormously increases the variation in ice-nucleation times – compare Figures 1 and 2 (the reason for which will be explained later) – but it also allows the small impurities (having a little effect on the ice-nucleation temperature $T_t$!) to either trigger or prevent the Mpemba effect. It's worth noting that the freezing temperature of even identically prepared and identically cooled pure water samples in identical test tubes varies by a fraction of a degree[37-40] between experiments.

According to the conventional Gibbs transition state theory for first-order phase transitions[36,41-43], like water-ice, the time of appearance, at a given temperature $T<T_0\equiv 273$ °K, of an ice nucleus (corresponding to the transition state, i.e., the most unstable state in ice formation) around one given H$_2$O molecule in pure, still water can be estimated as

$$t_T^{(1)} \sim \tau \cdot \exp\left(\frac{+G^{\#}_{\text{nucl}}(T)}{k_B T}\right). \qquad (4)$$

Here $G^{\#}_{\text{nucl}}(T)$ is the free energy of the ice nucleus at temperature $T$; $k_B$ is the Boltzmann constant; $\tau$ is the typical time for a "borderline" H$_2$O molecule to attach to ice; $\tau$ is no less than $\tau_0 \sim 10^{-12}$ s, the typical time[41] of thermal vibrations at 0 °C; in fact, $\tau$ is estimated (see[41], chs. 3.2, 8.2) from the difference between the rates of H$_2$O molecule attachment to and detachment from ice; $\tau$ depends on temperature much less than $\exp\left(\frac{+G^{\#}_{\text{nucl}}(T)}{k_B T}\right)$, which increases exponentially towards infinity as the temperature $T$ approaches 0 °C (so that, for kinetic reasons, the freezing temperature must be strictly below[36,37,41] 0 °C). Considering only the dominant terms[36,37,41], $\tau \sim 10^{-7}$ s at normal sub-zero temperatures of $-1°$ to $-20$ °C.

The free energy $G^{\#}_{\text{nucl}}$ is very large for a 3-dimensional ice nucleus in bulk water[36,37], where ice, for kinetic reason, arises only below[44] -35 °C. However, a 2-dimensional (2D) ice nucleus has a smaller $G^{\#}_{\text{nucl}}(T)$; thus, ice can arise[37] on an ice-binding border of water at negative temperatures close to 0 °C.

In the 2D case (and $T<T_0$), when the ice nucleus appears at the water's border, the free energy of this nucleus is estimated[36,37] as $G^{\#}_{\text{nucl}}(T) \approx 4B_2\left(\frac{B_2}{-\Delta\mu(T)}\right)$. Here $\Delta\mu(T)$ is the chemical potential of H$_2$O in ice minus that in water (so that $\Delta\mu(T_0)$=0, and $\Delta\mu(T)<0$ at $T<T_0$, where $\frac{\Delta\mu(T)}{k_B T} \approx \frac{T-T_0}{100°}$ at temperatures $T<T_0$ close to $T_0$, (see[37], with data taken from[45]), and $B_2$ is the additional free energy of an ice molecule at the ice/water interface. $B_3 \approx 0.85\, k_B T_0$ for a 3D ice/water interface (see[37], with data taken from[46]), which seems to remain more or less valid (i.e., $B_2 \approx k_B T_0$) for the perimeter of 2D ice layer[40].

With the above $\Delta\mu(T)$ and $B_2$ estimates, we have[37-40] that, at temperatures $T<T_0$ (and close to $T_0$),

$$\frac{G^{\#}_{\text{nucl}}(T)}{k_B T} = \frac{T_*}{T_0 - T}, \qquad (5)$$

where $T_* \approx 400°$ at $B_2 \approx k_B T_0$. The above estimate captures the main effect: an enormous growth of $G_{\text{nucl}}^{\#}(T)$ when $T$ approaches the ice-and-water coexistence temperature $T_0 \equiv 273$ °K (i.e., 0 °C).

Since the time of appearance, at a given temperature $T<T_0$, of an ice nucleus around *one* given on-border H$_2$O molecule is determined by Eqs. (4), (5), the time of appearance of an ice nucleus around *one of the N* on-border H$_2$O molecules (where ice can only appear) can be estimated as

$$t_T^{(N)} \sim \frac{\tau}{N} \cdot \exp\left(\frac{T_*}{T_0-T}\right). \tag{6}$$

With the number of on-border waters $N \sim 10^{15}$ in a ~1 mL test tube, and $\tau \sim 10^{-7}$ s, one can see from this equation (with $T_* \approx 400°$) that $t_T^{(N)}$ is minutes at $T_0 - T \approx 7°$, weeks at $T_0 - T \approx 6°$, ages at $T_0 - T \approx 5°$. Thus, experimentally, freezing initiation is only possible at temperatures below $\approx -6$°C.

The probability of appearance of an ice nucleus around one given on-border H$_2$O molecule during the above specified time $\tau$ at temperature $T<T_0$ is $p_{T<T_0,\tau}^{(+1)} \equiv \frac{\tau}{t_T^{(1)}} = \exp\left(\frac{-T_*}{T_0-T}\right)$, and $p_{T \geq T_0,\tau}^{(+1)} = 0$ at $T \geq T_0$. Thus,

$$p_{T<T_0,dt}^{(+1)} = \frac{dt}{\tau} \cdot \exp\left(\frac{-T_*}{T_0-T}\right) \tag{7}$$

is the probability of the nucleus appearance during a short time $dt$, and $p_{T,dt}^{(-1)} \equiv 1 - p_{T,dt}^{(+1)}$ is the probability that a nucleus *does not* appear during this time around this one on-border water molecule.

So, $p_{T<T_0,dt}^{(-N)} = \left[1-p_{T<T_0,dt}^{(+1)}\right]^N \approx \exp\left(-Np_{T<T_0,dt}^{(+1)}\right) = \exp\left[-dt\frac{N}{\tau}\exp\left(\frac{-T_*}{T_0-T}\right)\right]$ is the probability that, at $T<T_0$, in a system of $N$ on-border water molecules *not a single* nucleus arises during the time $dt$. Then $\prod_{i=1}^{m} p_{T_i<T_0,dt_i}^{(-N)} = \exp\left[-\sum_{i=1}^{m} dt_i \frac{N}{\tau}\exp\left(\frac{-T_*}{T_0-T_i}\right)\right]$ is the probability that *not a single* nucleus arises in this system during the time interval $t_0\text{—}t^* = dt_1 + \cdots + dt_m$ when all temperatures $T_1, \ldots, T_m$ are below $T_0$.

A freezer set at the temperature $T_{\text{th}}<T_0$ spends the time $t_0 = t(T_1, T_0)$, see Eq. (3), for cooling a sample from $T_1$ to $T_0$, after which ice can start to arise. At time above $t_0$, when $T_t<T_0$ (i.e., below 0 °C),

$$S_{t^*>t_0}^{-} = \exp\left[-\int_{t_0}^{t^*} dt\frac{N}{\tau}\exp\left(\frac{-T_*}{T_0-T_t}\right)\right] \tag{8}$$

is the probability that, at $T_t<T_0$, not a single nucleus arises among $N$ of on-border waters during time $t_0\text{—}t^*$ (note that $S_{t \leq t_0}^{-} = 1$, and $S_{t^*>t_0}^{-}$ tends to 0 at $t^* \to \infty$ with decreasing $T_{t^*}<T_0$). Then $S_{t \leq t_0}^{+} = 0$, and

$$S_{t^*>t_0}^{+} \equiv 1 - S_{t^*>t_0}^{-} = 1 - \exp\left[-\int_{t_0}^{t^*} dt\frac{N}{\tau}\exp\left(\frac{-T_*}{T_0-T_t}\right)\right] \tag{9}$$

is the probability that the first nucleus arises in the on-border water during time interval $t_0\text{—}t^*$.

The probability of ice nucleation during a short time interval $t^*\text{—}(t^*+dt^*)$ is

$$dt^* \cdot \left\{\frac{d}{dt^*}S_{t^*>t_0}^{+}\right\} = \left\{\exp\left[-\int_{t_0}^{t^*} dt\frac{N}{\tau}\exp\left(\frac{-T_*}{T_0-T_t}\right)\right] \cdot \left[\frac{N}{\tau}\exp\left(\frac{-T_*}{T_0-T_{t^*}}\right)\right]\right\} \cdot dt^*, \tag{10}$$

so that the density $P_{t^*}^{+}$ of the "accumulated" probability $S_{t^*}^{+}$ is 0 when $t^* \leq t_0$, and

$$P_{t^*>t_0}^{+} \equiv \frac{d}{dt^*}S_{t^*>t_0}^{+} = \exp\left[-\int_{t_0}^{t^*} dt\frac{N}{\tau}\exp\left(\frac{-T_*}{T_0-T_t}\right)\right] \cdot \left[\frac{N}{\tau}\exp\left(\frac{-T_*}{T_0-T_{t^*}}\right)\right]; \tag{11}$$

this $P_{t^*>t_0}^{+}$ is the momentary probability of ice nucleation at $t^* > t_0$ (note that $\int_{t_0}^{\infty} P_{t^*>t_0}^{+} dt^* = 1$, while the time needed to cool an ice-free sample to the temperature $T_{t^*}$ is given by Eq. (3).

Figure 3 shows the computed (according to Eq. (11) with $N=10^{15}$, $\tau = 10^{-7}$ s, $\kappa = 10^{-3}$ s$^{-1}$, $T_* = 400°$) probabilities $P_{t^*}^{+}$ of ice nucleation as functions of the time $t^*$ of cooling from the initial temperatures $T_1=+20$ and $T_1=+70$ °C at various thermostat temperatures $T_{\text{th}}$. It can be seen that the expected variations in freezing times for both these temperatures (Figs. 3**a**, 3**b**) are quite wide and overlap at higher freezer temperatures (above $-7$ °C), though at lower freezer temperatures these variations are narrow and do not overlap. This theoretical result agrees with experimental observations of identical water samples cooled at different freezer temperatures (Figs. 1, 2), supporting the conclusion that the "Mpemba effect"—in a fraction of experiments—manifests only at relatively high freezer temperatures $T_{\text{th}}$, not at lower ones.

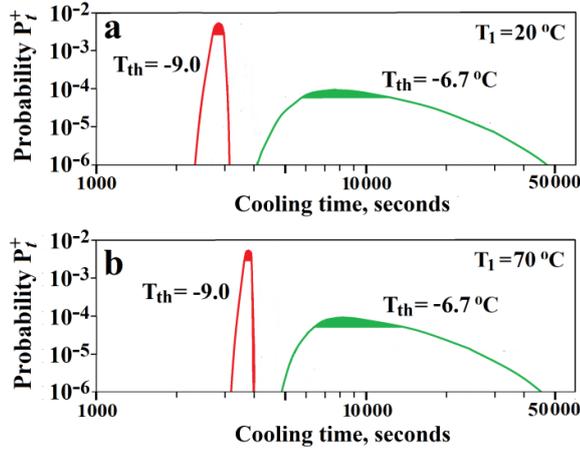

**Fig. 3. The probabilities $P_{t^*}^+$ of ice nucleation as functions of the cooling time at different freezer temperatures $T_{th}$.** The cooling begins at initial temperatures $T_1 = +20\,°C$ (**a**) and $T_1 = +70\,°C$ (**b**). The only difference between (**a**) and (**b**) is the time shift of about 1000 seconds in (**b**), reflecting the additional time required to cool hot water from $+70\,°C$ to $+20\,°C$. Notably, at higher $T_{th}$, the expected times of ice nucleation exhibit significant (>>1000 sec) variations. In contrast, at lower $T_{th}$ (below $-8\,°C$), the variations of freezing times are much smaller than 1000 sec. See colored caps on the tops of $P_{t^*}^+$ peaks.

To grasp why higher freezer temperatures $T_{th}$ result in broad freezing time ranges while lower yield narrow ranges, we will simplify the water cooling & freezing process into two stages: (a) *cooling*, where the water temperature decreases from $T_1$ to $T_{fr}$, and (b) *freezing onset* at the constant temperature $T_{fr}$.

At cooling by a freezer with temperature $T_{th}$, the most probable ice-nucleation temperature $T_{fr}(T_{th})$ corresponds to the minimum (by $T_{fr}$) of the sum of the (a) cooling time, $t_{cooling}(T_1, T_{fr}(T_{th})) = \frac{1}{\kappa} \cdot \ln\frac{T_1 - T_{th}}{T_{fr} - T_{th}}$ (see Eq. (3)), and (b) the following (after reaching $T_{fr}$) mean freezing onset time $\langle t \rangle_{T_{fr}=const}^{freezing} = \int_0^\infty t^* \cdot P_{t^*}^+ \cdot dt^*$. At the constant $T_t = T_{fr}$, $P_{t^*}^+ = \frac{d}{dt^*}\left[1 - \exp\left[-(t^* - 0)\frac{N}{\tau}\exp\left(\frac{-T_*}{T_0 - T_{fr}}\right)\right]\right]$ according to Eq. (11); denoting the constant value of $\frac{N}{\tau}\exp\left(\frac{-T_*}{T_0 - T_{fr}}\right)$ as $A_{T_{fr}}$, we have $P_{t^*}^+ = A_{T_{fr}} e^{-A_{T_{fr}} t^*}$, and

$$\langle t \rangle_{T_{fr}}^{freezing} = \int_0^\infty t^* \cdot P_{t^*}^+ \cdot dt^* = \int_0^\infty t^* \cdot A_{T_{fr}} e^{-A_{T_{fr}} t^*} \cdot dt^* = \frac{1}{A_{T_{fr}}} \equiv \frac{\tau}{N} \exp\left(\frac{T_*}{T_0 - T_{fr}}\right), \quad (12)$$

which drastically, exponentially grows when $T_{fr}$ approaches $T_0$.

The above mentioned minimum (by $T_{fr}$) of the sum $\frac{1}{\kappa} \cdot \ln\frac{T_1 - T_{th}}{T_{fr} - T_{th}} + \frac{\tau}{N}\exp\left(\frac{T_*}{T_0 - T_{fr}}\right)$ is achieved when

$$\frac{[(T_0 - T_{fr})/T_*]^2}{[(T_0 - T_{th})/T_*] - [(T_0 - T_{fr})/T_*]} = \frac{\tau}{N}\exp\left(\frac{T_*}{T_0 - T_{fr}}\right)\frac{1}{1/\kappa}. \quad (13)$$

Figure 4**a** presents the resulting $T_{fr}(T_{th}) - T_0$ as a function of $T_{th} - T_0$.

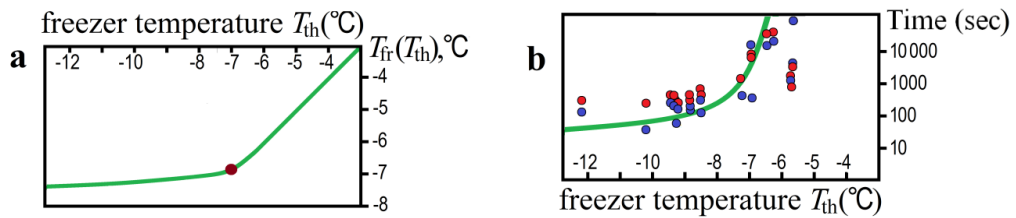

**Fig. 4. Characteristics of the "cooling & freezing" process as a function of the freezer temperature $T_{th}$.** (**a**) Ice nucleation temperature $T_{fr}(T_{th})$ according to Eq. (13). (**b**) Experimentally found ranges of freeing times at various $T_{th}$ values (blue and red dots – for the initially cold, $+20\,°C$, and hot, $+70\,°C$, water samples), and the theoretically expected root mean square deviation $RMSD_{T_{fr}}^t$ of the experimental freezing times from their means at given $T_{fr}(T_{th})$ values (green line – see Eq. (15) below). At $T_{th} > -7\,°C$, the range of the freeing onset times exceeds the cooling lag of the initially hot water, which is ≈500-1000 sec (see Figs. 1, 2), and the initially hot water may stochastically freeze before the cold water.

The solution of Eq. (13) has two regimes: (A) when $\frac{[(T_0-T_{\text{fr}})/T_*]^2}{(T_{\text{fr}}-T_{\text{th}})/T_*} > 1$, and (B) when $\frac{[(T_0-T_{\text{fr}})/T_*]^2}{(T_{\text{fr}}-T_{\text{th}})/T_*} < 1$; they are separated by the condition $\frac{[(T_0-T_{\text{fr}})/T_*]^2}{(T_{\text{fr}}-T_{\text{th}})/T_*} = \frac{\kappa\tau}{N}\exp\left(\frac{T_*}{T_0-T_{\text{fr}}}\right) = 1$, defining the break point (brown in Fig. 4a) of the plot: here, $T_{\text{fr}} - T_0 = -T_*/\ln\left(\frac{N}{\kappa\tau}\right)$, and $T_{\text{th}} = T_{\text{fr}} - T_*/\left[\ln\left(\frac{N}{\kappa\tau}\right)\right]^2$. With the above values of $N, \tau, \kappa, T_*$: $T_{\text{fr}} \approx -6.9$ °C, $T_{\text{th}} \approx -7.1$ °C.

Having found the dependence of $T_{\text{fr}}$ on $T_{\text{th}}$, we can find, see Eqs. (3, 12), the $T_1 \to T_{\text{fr}}$ cooling time $t_{\text{cooling}}(T_1, T_{\text{fr}}(T_{\text{th}}))$ and the mean time of the subsequent freezing onset $\langle t\rangle^{\text{freezing}}_{T_{\text{fr}}=\text{const}}$ as functions of $T_{\text{th}}$.

In the case (A), when $\frac{[(T_0-T_{\text{fr}})/T_*]^2}{(T_{\text{fr}}-T_{\text{th}})/T_*} > 1$ (since, when $T_{\text{th}} > -7.1$ °C, $T_{\text{fr}} - T_{\text{th}}$ is *small*), $\frac{\tau}{N}\exp\left(\frac{T_*}{T_0-T_{\text{fr}}}\right)$ exceeds $1/\kappa$ and exponentially grows with decreasing $T_0 - T_{\text{fr}}$ (which, when $T_{\text{fr}} - T_{\text{th}}$ is small, is very close to $T_0 - T_{\text{th}}$). In this case, the freezing onset time $\langle t\rangle^{\text{freezing}}_{T_{\text{fr}}=\text{const}}$ can be exponentially large.

In the case (B), when $\frac{[(T_0-T_{\text{fr}})/T_*]^2}{(T_{\text{fr}}-T_{\text{th}})/T_*} < 1$ because $T_{\text{fr}} - T_{\text{th}}$ is large, $\frac{\tau}{N}\exp\left(\frac{T_*}{T_0-T_{\text{fr}}}\right)$ is less than $1/\kappa$, and the freezing onset time $\langle t\rangle^{\text{freezing}}_{T_{\text{fr}}=\text{const}}$ is relatively small.

Despite some stochastic deviations of local temperatures from their average values[47], for $t_{\text{cooling}}$ they cannot result in significant stochastic variances, because the $t_{\text{cooling}}$ dependence on $T_{\text{fr}} - T_{\text{th}}$ is only logarithmic, i.e. weak. But the variances in the freezing onset times $\langle t\rangle^{\text{freezing}}_{T_{\text{fr}}}$ are really large and important. They are estimated as follows.

As known, the expected root mean square deviation ($\text{RMSD}_X$) of experimental $X$ values from their mean $\langle X\rangle$ is $\sqrt{\langle X^2\rangle - \langle X\rangle^2}$, where $\langle X^2\rangle$ is the mean square of $X$. The mean square for the freezing time is

$$\langle t^2\rangle^{\text{freezing}}_{T_{\text{fr}}=\text{const}} = \int_0^\infty (t^*)^2 \cdot P^+_{t^*} \cdot dt^* = \int_0^\infty (t^*)^2 \cdot dt^* \cdot A_{T_{\text{fr}}} e^{-A_{T_{\text{fr}}}t^*} = \frac{2}{A^2_{T_{\text{fr}}}} \equiv 2\left[\frac{\tau}{N}\exp\left(\frac{T_*}{T_0-T_{\text{fr}}}\right)\right]^2, \quad (14)$$

so that the expected root mean square deviation of the experimental freezing times from their mean is

$$\text{RMSD}^t_{T_{\text{fr}}=\text{const}} = \sqrt{\langle t^2\rangle_{T_{\text{fr}}=\text{const}} - \langle t\rangle^2_{T_{\text{fr}}=\text{const}}} = \frac{\tau}{N}\exp\left(\frac{T_*}{T_0-T_{\text{fr}}}\right), \quad (15)$$

which drastically, exponentially grows when $T_{\text{th}}$ (and hence $T_{\text{fr}}$) approach 0 °C – see Fig. 4.

Thus, the resulting estimate of the time of the whole cooling & freezing process is

$$\left[t_{\text{cooling}}(T_{\text{fr}}) + \langle t\rangle^{\text{freezing}}_{T_{\text{fr}}=\text{const}}\right] \pm \text{RMSD}^t_{T_{\text{fr}}=\text{const}} = \left[\frac{1}{\kappa}\cdot\ln\frac{T_1-T_{\text{th}}}{T_{\text{fr}}-T_{\text{th}}} + \frac{\tau}{N}\exp\left(\frac{T_*}{T_0-T_{\text{fr}}}\right)\right] \pm \frac{\tau}{N}\exp\left(\frac{T_*}{T_0-T_{\text{fr}}}\right), \quad (16)$$

where the exponential term is dominating when $T_{\text{th}}$ (and hence $T_{\text{fr}}$) approach 0 °C.

Thus, we theoretically expect (and see in experimental Figs 1, 2) long, highly variable ice-nucleation times when $T_{\text{th}}$ is above $-7$ °C, and short, low- variable ice-nucleation times when $T_{\text{th}}$ is below $-8$ °C

The variability of ice-nucleation times is typical for first-order phase transitions. This large variability is due to the specific form of the time-dependent probability of such transitions, $P^+_{t^*} \sim e^{-A_{T_{\text{fr}}}t^*}$, which exhibits its maximum at the first moment and then declines exponentially. This leads to substantial uncertainty in the timing of stochastic ice nucleation and, at sufficiently high $T_{\text{th}}$, results in overlapping freezing times for samples with different initial temperatures, providing the stochastic Mpemba effect.

It is important to emphasize that the observed stochasticity of the Mpemba effect is limited to phenomena such as freezing, which involve first-order phase transitions.

The initiation of freezing is, in essence, "a single-event phenomenon": the emergence of a single ice nucleus can trigger the freezing of the entire liquid. Therefore, the significant uncertainty in the timing of stochastic ice nucleation can, in *some* experiments, *occasionally* demonstrate the Mpemba effect. This occurs when the broad ranges of ice initiation times for hot and cold samples overlap, potentially masking the delay in cooling of the hotter water sample and leading to the perception that *"hot water* [deterministically] *freezes faster than cold water."*

We do not want to create the impression that our finding of the stochasticity of the Mpemba effect in freezing disproves the very existence of such effects. A simple mechanical system[35] provides an example of a strict Mpemba effect, where "*the state that is initially farthest from its equilibrium state attains the latter at the earliest time*", and where the origin and driving force of this effect are crystal clear.

The observed stochasticity of the Mpemba effect in freezing of pure water does not exclude the possibility that, in some cases—when vessel walls (or liquid) are not clean enough (we used measures to minimize wall effects, see Methods), the Mpemba effect, not based on the properties of the pure liquid *itself*, can result from foreign ice nucleators entering the hot liquid. In particular, natural water is often contaminated with *P. syringae* bacteria, which are potent ice nucleators that, at a concentration of only 0.01 mg/ml, can raise the freezing point[40] by ≈1°. The resulting, possibly small, rise in freezing temperature can lead—see Eq. (16)—to a significant acceleration of ice nucleation in the hot liquid (and thus, the Mpemba effect) when the freezer temperature $T_{th}$ is close to the freezing point of the liquid.

**METHODS**

The experiments were conducted in the basement of a separate building, where all vibration sources were turned off to prevent the impact on ice freezing. A Tuvio FCS15HV1 chest freezer was used; its horizontal lid minimizes temperature fluctuations at its opening – as compared to freezers with a vertical door. This freezer allowed the temperature to be set in the range from +13 °C to -30 °C with an accuracy of ±0.5 °C. We usually set -14 °C; this is a temperature at the bottom of the chest freezer. The freezer (except for the heat exchanger) was well insulated. To further stabilize the temperature, about 30 kg of frozen water in big plastic bottles were placed in the freezer.

Between these bottles we installed an "inner chamber" – a foam plastic box with 11 small plastic bottles containing 310 ml of water with 15% NaCl, which were cooled to -12 °C before each experiment. The frozen salt water thawed slowly, which allowed us to reduce the temperature fluctuations inside the "inner chamber" to ±0.5 °C, while outside this chamber, turning the compressor on and off was accompanied by a temperature change of ±2 °C. We started the experiment when the temperature in the inner chamber near the small bottles with salt water (measured by a separate thermistor, i.e. a temperature-sensitive resistor located at the end of a long wire connected to an external digital thermometer) reached the desired value for this experiment.

Inside the inner chamber, a lightweight wire stand was suspended using 18 thin, 100 cm long wires. Nine thermistors, connected to external electronics, were attached to the ends of each pair of these wires. Eight of these thermistors were inserted in the eight closed test vials through holes in the vial lids. Thus, each of these test vials (each with a capacity of 2 ml) was suspended on a pair of wires (minimizing its potential vibrations), with a thermistor at the end of this pair. Water occupied ≈80% of the vial volume, and the thermistor measured temperature in its upper half. The vials and water in them were weighed on scales with an accuracy of 10 mg. The vials with four initially hot (≈+70 °C) and four initially cold (≈+20 °C) water samples were positioned in the wire stand in an alternating pattern, with a 4 cm separation. They were placed into the freezer all together. The ninth thermistor was in air, between the small bottles with salt water and test vials.

Nine pre-calibrated (using the digital thermometer LT-300 Termexlab.ru with a resolution of 0.01°C) MF59 100K thermistors, each with a 1-meter thermowire, were used to measure the water temperature within the test vials and the air temperature near the small bottles with salt water in the inner chamber. Three pre-calibrated Boomshakalaka digital thermometers (resolution of 0.1 °C) with sensors at the ends of 1-meter wires were additionally used to monitor temperature inside the inner chamber, the ambient air, and the electrical heater used to preheat the water samples.

We used a custom-built data acquisition module that included a computer running specialized software, which sent 5V pulses sequentially to each of the nine thermistors via the LPT1 parallel port and a custom adapter. Digital voltage readings, representing temperature, were collected from the Radio Shack 22-168A digital multimeter connected to the RS232 serial port. This multimeter measured the voltage across a single 100 kΩ resistor (shared by all thermistors and connected to ground), the current through which depended on the temperature of the thermistor.

Temperature measurements and corresponding timestamps were recorded every 43 seconds.

The thermistors in the test vials and interiors of the vials were lubricated with "Sylgard 184 silicone elastomer" (without the "Curing Agent") to adsorb potential foreign ice nucleators from water and to smooth out any micro-roughness on the vial walls.

Upon inserting the stand with vials, a slight warming of the air in the upper part of the inner chamber was observed. However, the air temperature, continuously monitored by a digital thermometer and the thermistor near the small bottles with salt water and vials returned to the set temperature within 10 minutes.

Distilled water samples were used in all experiments.

Before each experiment, four covered vials were equilibrated at the ambient basement air temperature (+15 – +20 °C), and four covered vials, placed in four small homemade electric ovens, were pre-heated to ≈+70 °C.

**Acknowledgments:** We are grateful to Bogdan S. Melnik for discussions and assistance in experimental and theoretical work and E. V. Serebrova for editing the manuscript.

**Author Contributions:** Conceptualization, A. V. F.; Literature collection and analysis, A. V. F. & A. A. K.; Experiments, A. A K.; Writing - Original Draft Preparation, Review & Editing, A. V. F.

**Funding:** This work has been supported by the Russian Science Foundation (grant No. 25-14-00260).

**Competing interests** The authors declare no competing interests.

**Correspondence and requests for materials** should be addressed to A. V. F. (afinkel@vega.protres.ru) and A. A. K. (anklimov11@gmail.com).